\documentclass{aip-cp}  
\usepackage[numbers]{natbib}
\usepackage{rotating}
\usepackage{graphicx}
\usepackage{epsfig}
\usepackage{psfrag}

\def\auj{\number\day\space\ifcase\month\or
janvier\or f\' evrier\or mars\or avril
\or mai\or juin\orjuillet\or ao\^ut
\or septembre\or octobre\or novembre
\or d\' ecembre\fi\space\number\year}

\def\hoje{\number\day\space de \ifcase\month\or
Janeiro,\or Fevereiro,\or Mar\cc o,\or 
Abril,\or Maio,\or Junho,\or Julho,
\or Agosto,\or Setembro,\or Outubro,\or Novembro,
\or Dezembro,\fi\space\number\year}

\def\today{\number\day\space de \ifcase\month\or
January,\or Febreary,\or March,\or 
April,\or May,\or June,\or July,
\or Agust,\or September,\or October,\or November,
\or December,\fi\space\number\year}

\newcommand{\be}{\begin{equation}}
\newcommand{\ee}{\end{equation}}
\newcommand{\bea}{\begin{eqnarray}}
\newcommand{\eea}{\end{eqnarray}}
\newcommand{\bno}{\begin{eqnarray*}}
\newcommand{\eno}{\end{eqnarray*}}

\newcommand{\cc}{\c c}

\newcommand{\bl}{\begin{large}}
\newcommand{\el}{\end{large}}
\newcommand{\bla}{\begin{Large}}
\newcommand{\ela}{\end{Large}}

\newcommand {\sla} {\slash \hspace{-0.18cm} }     
\newcommand{\hsp}{\hspace{0.70cm}}  

\newcommand{\ede}{\end{document}}  

\begin{document}

\title{ Parton Distribution in Pseudoscalar Mesons with a Light-Front Constituent Quark Model     } 
\author[aff1]{J.~P.~B.~C.~de~Melo
\corref{cor1}}
\author[aff1,aff2]{Isthiaq Ahmed}
\author[aff1]{Kazuo Tsushima}
\eaddress{kazuo.tsushima@cruzeirodosul.edu.br}
\affil[aff1]{Laborat\'orio de F\'\i sica Te\'orica e Computacional\\
Universidade Cruzeiro do Sul\\
01506-000, S\~ao Paulo, SP, Brazil}
\affil[aff2]{  National Center for Physics, Quaidi-i-Azam University Campus, 
Islambad, 45320 Pakistan
}
\corresp[cor1]{Corresponding author: 
joao.mello@cruzeirodosul.edu.br } 
\maketitle

\begin{abstract}
We compute the distribution amplitudes of the pion and kaon in the light-front 
constituent quark model with the symmetric 
quark-bound state vertex function~\cite{deMelo2002,deMelo2014,Yabusaki2015}. 
In the calculation we explicitly include the flavor-SU(3) symmetry breaking effect 
in terms of the constituent quark masses of the up (down) and strange quarks. 
To calculate the kaon parton distribution functions~(PDFs), 
we use both the conditions in the light-cone wave function, i.e.,  when $\bar{s}$ quark is 
on-shell, and when $u$ quark is on-shell, and make a comparison between them. 
The kaon PDFs calculated in the two different conditions clearly show asymmetric behaviour 
due to the flavor SU(3)-symmetry breaking implemented by the quark masses~\cite{Nam2012v1,Nam2012v2}. 
\end{abstract}

\section{INTRODUCTION}

\hsp
In the usual hadronic scale the pion is special, because it is (unusually) the 
lightest bound state of the quark-antiquark, and also plays a key role 
in the chiral perturbation theory~(see the review~\cite{Bira1999}).

Deep inelastic scattering~(DIS), or the hadronic scattering, 
may be decomposed into two parts, a part depends on the short distance partonic 
cross section, and a part depends on the long distance 
process~\cite{Nam2012v1,Nam2012v2,Frederico1994}. 
The distribution amplitude~(DA) provides the link between these two parts, 
or, with another words, pertubative regime~(small distance) and 
non-pertubative~(long distance) regime.

The hadronic states, mesons and baryons, have been sucessfully 
organized in terms of the valence quark components, or constituent picture 
(see the Particle Data Group~\cite{PDG}). 
However, on the other hand, in the case of the {\it Deep Inelastic Scattering}, 
the hadronic systems also have other degrees of freedom, 
i.e., gluons and the sea of quarks. In the hadronic systems, the 
distributions amplitudes (DAs)~are important to characterize the feature of the 
bound state nature, because they provide essential information  
on them, such as electromagnetic and transition form factors~\cite{Choi2007,Nico2015}.

In the following sections we present the model utilized, and the results.

\section{The Model}

\hsp
The model used in the present study is the light-front constituent quark model~(LFCQM) 
with a symmetric vertex function, which was used already in 
the previus work for pion~\cite{deMelo2002,deMelo1999}, 
and recently for pion and kaon~\cite{Yabusaki2015} 
to calculated the observables associated with them. The effective interaction 
Lagrangian density for the quark and pseudoscalar meson is given by, 

\begin{eqnarray}
  \mathcal{L}_\mathrm{eff} & = &  -ig\, \vec\phi\!\cdot\bar q\gamma^5\vec\tau q \ , 
\end{eqnarray}
where $g=m_{0^-}/f_{0^-}$ 
is the coupling constant, $m_{0^-}$ and $f_{0^-}$ denote the mass 
and decay constant of a pseudoscalar 
meson, respectively, and $\vec \phi$ 
the pseudoscalar field, namely the pion or the 
kaon~\cite{deMelo2002,Yabusaki2015}. Here,   
the symmetric vertex function for the $\bar{q}q$-meson vertex 
which describes the bound state, is used as in  
Refs.~\cite{deMelo2002,Yabusaki2015}:

\begin{eqnarray}
\Lambda (k,P) ~=  ~\mathcal{C} \big [ \frac{1}{(k^2-m_R^2 + i\epsilon)} 
                        +   \frac{1}{((P-k)^2-m_R^2+i\epsilon)} \big ] . 
\label{vertex}
\end{eqnarray}
The vertex function $\Lambda(k,p)$ is symmetric under the exchange of the quark 
and antiquark momenta, $k$ and $P-k$, with $P$ being the total momentum of the 
meson. The normalization constant,
$C$, is fixed by the electromagnetic charge form factor, 
or the charge i.e., $F_{0^-}(0)=1$.

The pseudoscalar meson wave function either for the pion or kaon in the present 
light-front constituent quark model, is given 
by~\cite{deMelo2002,Yabusaki2015}
\begin{eqnarray}
 \psi(k,p)_{\pi,K} = \frac{m_q}{f_{\pi,K}}\frac{\sla{k}+m_q}{k^2-m_q^2+i\epsilon}
 \gamma^5\Lambda(k,p)
 \frac{\sla{k}-\sla{p} + m_{\bar q}}{(k-p)^2-m_{\bar q}^2+i\epsilon} 
~ + ~[q \leftrightarrow\bar q].
\end{eqnarray}
In the expression above, the momentum~$k$ for the quark~(antiquark) is 
on-shell, namely, ~$k^-=\frac{k^2_{\perp}+m_{q\,(\bar{q})}^2}{k^+}$, and after 
the integration of the light-front energy,~$k^-$, the 
pseudoscalar valence light-cone wavefunction is obtained as,   
\begin{eqnarray}
\Phi(x,k_\perp,p^+,\vec{p}_{\perp}) =  
\bigg [ \frac{1}{(1-x)(m_{0^-}^2-\mathcal{M}^2(m_q^2,m_R^2))} 
 + \frac{1}{x(m_{0^-}^2-\mathcal{M}^2(m_R^2,m_{\bar q}^2))} \bigg ] 
 \frac{1}{m_{0^-}^2-\mathcal{M}^2(m_q^2,m_{\bar q}^2)}
  + [q~  \leftrightarrow ~\bar{q} ], 
  \label{wfunction}
\end{eqnarray}
with the definition for the functions~$\mathcal{M}(m_a,m_b)$, 
\begin{eqnarray*}
 \mathcal{M}^2(m_a^2,m_b^2)=
 \frac{k_\perp^2+m_a^2}{x}+\frac{(p-k)_\perp^2+m_b^2}{(1-x)}-p_\perp^2,
 ~ \nonumber
\end{eqnarray*}
where the ligh-front coordinates 
are,~$p^{\mu}=(p^+,p^-,\vec{p}_{\perp})=(p^0+p^3,p^0-p^3,\vec{p}_{\perp})$ and  
$x=k^+/p^+=(k^0+k^3)/(p^0+p^3)$. The second term in Eq.~(\ref{wfunction}) arises, 
because the vertex function used here is symmetric under 
the exchange of the quark and anti-quark momenta. But this is not the case of 
the non-symmetric vertex~\cite{deMelo1999,Otoniel2012}.

In the light-front the Feymman amplitude,  as calculated here, in the 
Breit frame and the Drell-Yan conditions~$q^+=0$, does not have zero-mode 
contribuitions. But in a general frame, i.e,~$q^+\neq 0$,~ it is necessary 
to include the zero-mode contribuitions~\cite{deMelo2002,deMelo1999,Bakker2001}. 
This is also true in the case of the other components, such as the minus component  
of the current~\cite{deMelo1999,Bakker2001} to 
preserve the full covariance. 

In general, the parton distribution function (PDF) of either pion or kaon is 
related with the corresponding wave function by the following equation,
\begin{equation}
 \phi_{\pi,K}(x)=
 \frac{2\sqrt{6}}{f_{\pi,K}} \int 
 \frac{dk^2_{\perp}}{16\pi^3}\Phi_{\pi,k}(x,k_\perp),
 \label{pdf}
\end{equation}
here, $f_{\pi,K}$ is the decay constant of pion or kaon, 
and $\Phi_{\pi,K}(x,k_\perp)$ the corresponding 
valence wave function. 

The pion or kaon wave function is normalized by 

\begin{equation}
 \int_0^1 dx \int \frac{d^2 k_\perp}{16 \pi^3 }
 \Phi_{\pi,k}(x,k_\perp)~=~ \frac{f_{\pi,K}}{2 \sqrt{6}},
\end{equation}
with the corresponding pseudoscalar meson decay constant.

In the present light-front constituent quark model (LFCQM) we have three parameters, 
the quark and antiquark masses and the 
regulator mass. With the constraint of the electromagnetic charge 
form factor,~$F_{\pi,K}(0)=1$, and 
the equation above for the normalization of the wave function, the parameters are 
determined. The LFCQM with the symmetric vertex function, has been successful in 
describing the pion and kaon observables with the data in the particle 
data group~(PDG)~\cite{PDG}~(see table-1).

\begin{center}
\begin{tabular}{|l|l|l|l|l|l|l|}
\hline  
\hline
\multicolumn{7}{|l|}{Table-1:~Observables for the pion and kaon} \\
\hline
          & $f_{0^-}$~(MeV) & $r_{0^-}$  & $m_u~(\pi^+)$ & $m_{\bar{d}}~(\pi^+)$  & $m_u~(K^+)$ 
          & $m_{\bar{s}}~(K^+)$ \\
  \hline 
Pion      & 93.12     &   0.736   & 220         &   220    &           &             \\
          & 101.85    &   0.670   & 250         &    250   &           &              \\  
\hline 
Kaon      & 101.81     &  0.754     &          &           & 220        &  440    \\
          & 113.74     &  0.687     &          &       & 250        &    440    \\       
\hline
\hline   
 \multicolumn{7}{|l|}{ $m_R~=~600$~MeV,~(all masses in MeV and radii in fm )  } \\
 \multicolumn{7}{|l|}{ Ex.(Pion): $f_{\pi}=92.4\pm0.021$~MeV, 
 $r_{\pi}=0.672\pm0.08$~fm~(PDG)~\cite{PDG}}\\
 \multicolumn{7}{|l|}{ Ex.(Kaon): $f_{k^+}=110.38\pm0.1413$~MeV, 
 $r_{k^+}=0.560\pm0.031$~fm~(PDG)~\cite{PDG}}\\
 \hline
 \hline
\end{tabular}
\label{table1}
 \end{center}

In the next section, we will present the numerical results 
for the parton distribution functions of the pion and kaon.

\section{Results and discussions}

\hsp
The parameters for the present model of the light quarks are the same as those in   
Refs.~\cite{deMelo2002,Yabusaki2015,deMelo2014}, and 
in the case of the strange quark, the strange quark mass,~$m_s (=m_{\bar{s}})$,~is taken from 
Ref.~\cite{Suisso2002}. In the previus works~\cite{deMelo2002,Yabusaki2015},  
the value of the regulador mass,
~$m_R=0.6$~GeV, was utilized to reproduce the pion and kaon observables~(see table-I) 
such as, the electromagnetic elastic form factors, decay constants and electromagnetic radii.

We compare the present results calculated with the symmetric vertex function 
with those of the non-symmetric vertex model~\cite{deMelo1999,Otoniel2012}, 
e.g., the ratios of the decay constants for the pion and 
kaon. While the ratio for the non-symmetruic vertex mofel gives $f^{nsy}_K/f^{nsy}_{\pi} \approx 1.363$, 
in the case of the present symmetric vertex model, we have the ratio,~$1.189$, which is 
very close to the experimental value from PDG~\cite{PDG}~$1.197\pm0.002\pm0.006$.

We now apply the model with the same set of parameters utilized in the previus 
works, in order to calculate the distribution amplitudes 
with~{\it LFCQM} model, and together with Eq.~(\ref{pdf}). 

\begin{figure*}[tbh]
\vspace{0.8cm}
\begin{center}
\includegraphics[scale=.30]{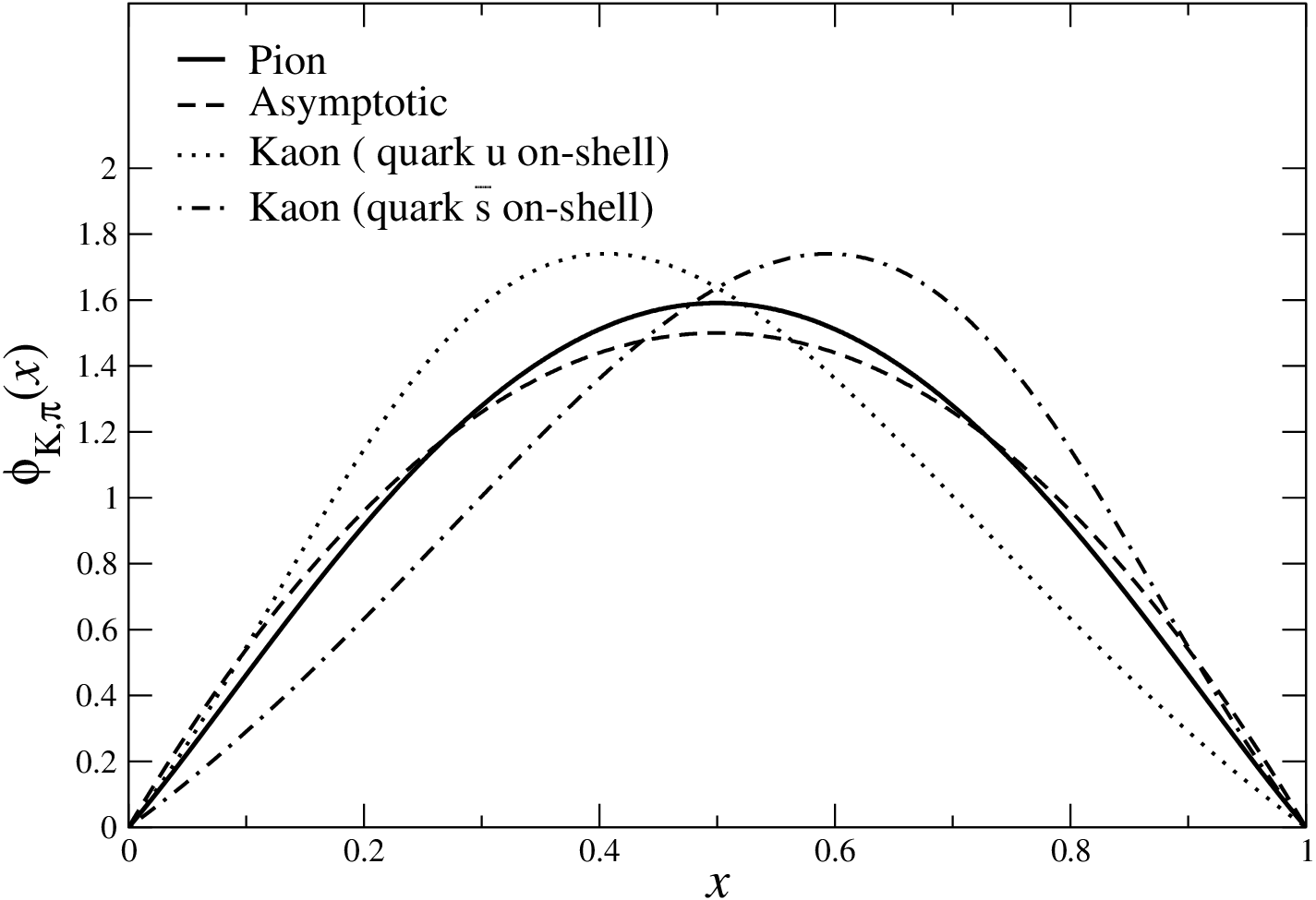} 
\hspace{0.20cm}
\includegraphics[scale=0.30]{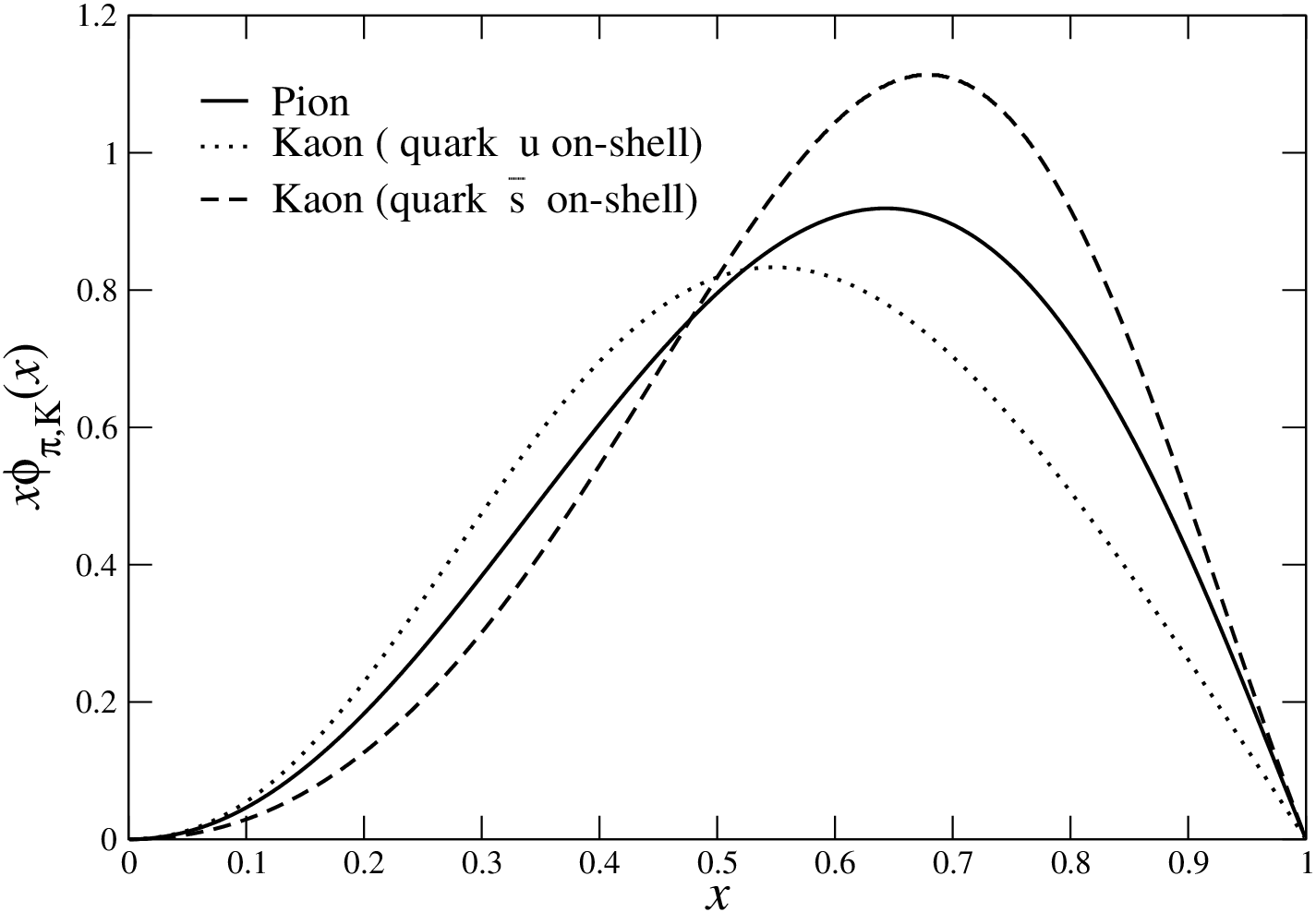}
\caption{(a)~(Left)~Pion and Kaon distribution amplitudes 
~(DAs) with the LFCQM with symmetric vertex~\cite{deMelo2002},  
and the asymptotic wave function~\cite{Lepage1980}, 
and (b)~(Right), the DAs of pion and kaon multiplied by x~($x\phi(x)$).
\label{fig1}
}
\end{center}
\end{figure*}

The results are shown in Figs.~1 and 2 for the distribution amplitudes~(DAs). 
In Fig.~1, it is shown the pion DA calculated with the present symmetric vertex model, 
compared with the pion asymptotic wave function~\cite{Lepage1980}; 
the result for the model presented here, is very similar to that of 
the pion asymptotic wave function. The pion DA (see the right panel of Fig.~\ref{fig1}), 
is symmetric around $x=0.5$ for both cases, the asymptotic wave function and the 
present model.

However, in the case of
the kaon, the DA is not symmetric for the interchange of the quark~($u$) and the 
antiquark~($\bar{s}$), that can be seen in the left panel of Fig.~\ref{fig1}. 
This is due to the choice of the integration momentum of the  
(anti)particle , i.e., with the $u$ quark on-shell, or the $\bar{s}$ quark on-shell~($\bar{s}$). 
This is the effect of the $SU(3)$ symmetry breaking
induced by the constituent quark masses considered in this study. 
Similar results were also obtained in  
Refs.~\cite{Nam2012v1,Nam2012v2} with a nonlocal chiral model.

\begin{figure*}[tbh]
\vspace{.8cm}
\begin{center}
\includegraphics[scale=.3]{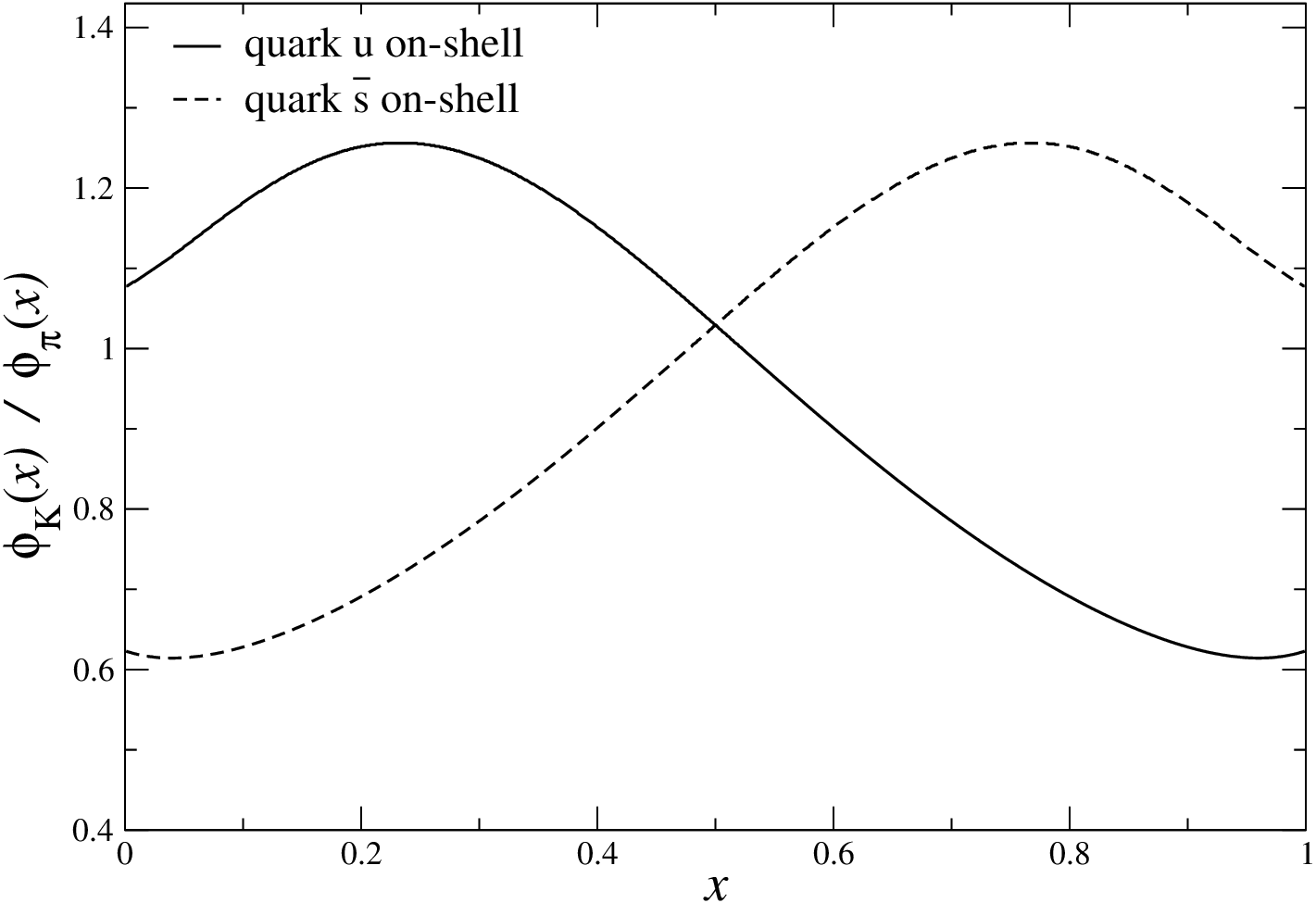} 
\hspace{0.20cm}
\includegraphics[scale=0.3]{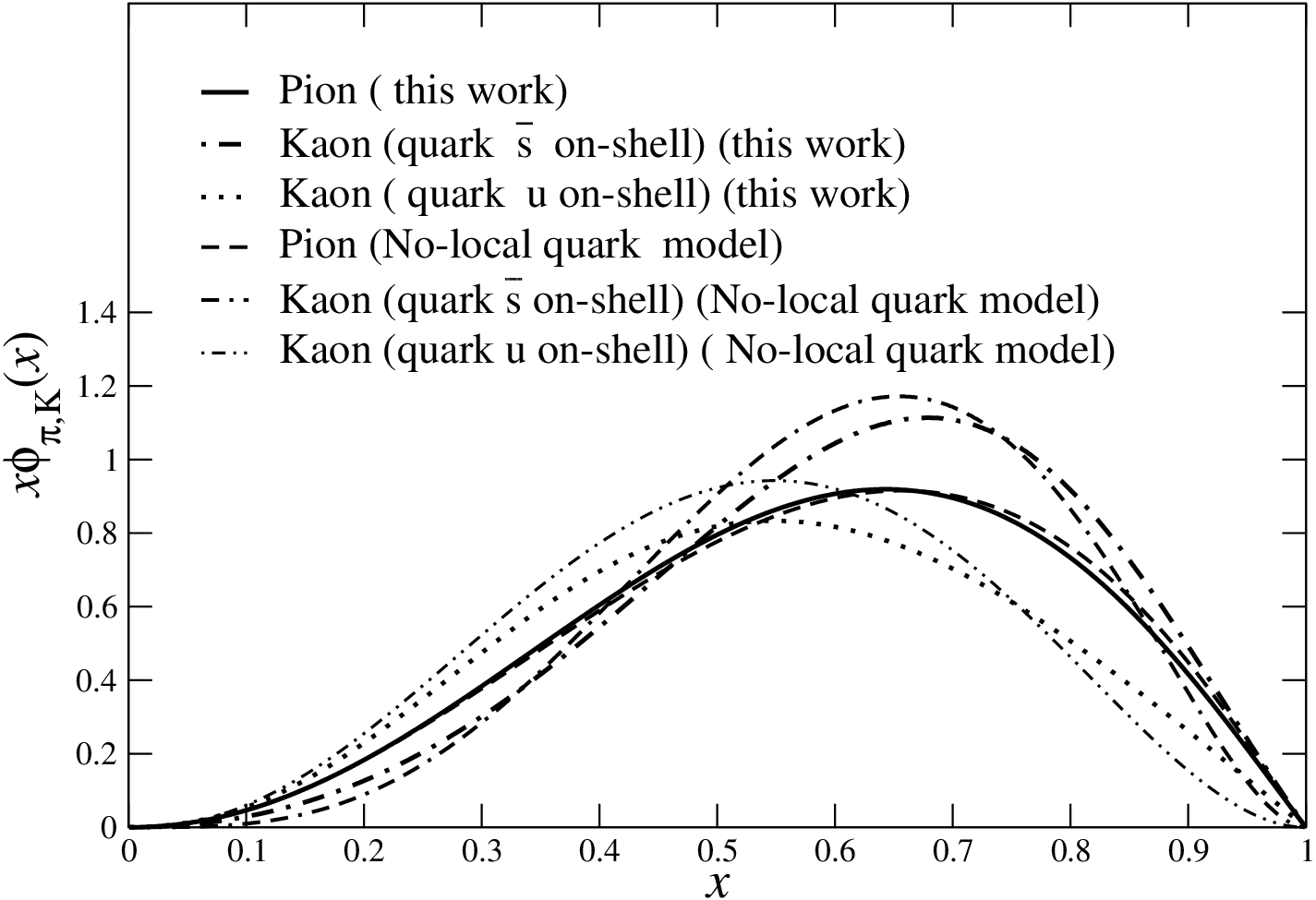}
\caption{(Left)~The ratios for the parton distribution functions $\phi(x)$ for the 
pion and kaon.
\newline 
(Right),~Parton distribution functions (PDFs) $\phi(x)$ for the 
pion and kaon multiplied by $x$,~compared with nonlocal 
chiral model~\cite{Nam2012v1,Nam2012v2}}.
\label{fig2}
\end{center}
\end{figure*}

In Fig.~\ref{fig2} (the left panel) we show the ratios of the 
distribution amplitude,~$u_K(x)/u_{\pi}(x)$, the kaon over pion with the 
quark~$u$ and antiquark on-shell ($\bar{s}$ quark on-shell in the case of the kaon). 
Again, the effect of the flavor-SU(3) symmetry breaking is clearly seen by the curves; 
the solid curve is the quark~$u$ on-shell, while the dashed curve is the case 
of the anti-strange quark~$\bar{s}$ on-shell. We show in the right panel of  
Fig.~\ref{fig2} the~PDFs for the pion and kaon multiplied by $x$, 
compared with the nonlocal quark model of Refs.~\cite{Nam2012v1,Nam2012v2}.

In the present work, the main objective is to study the 
behaviour of DAs and PDFs with the LFCQM of Ref.~\cite{deMelo2002}, 
since that model descrives very well the pion observables as well as those of the 
kaon~\cite{Yabusaki2015}. For the case of the kaon in the present study, both the 
DAs and PDFs show asymmetric behavior, similar to those found 
in different models. 

In the near fututure we plan to study the pion and kaon DAs and PDFs in the 
nuclear medium as the extension of the study made in Ref.~\cite{deMelo2014}.
Such studies are under in process.

\vspace{1.cm}
\noindent
{\bf Acknowledgement}\\
This work was partially supported by the Brazilian agencies CNPq, FAPESP 
and Universidade Cruzeiro do Sul (UNICSUL). 
The authors thank the organizers of XVI International Conference on Hadron Spectroscopy for
the invitations, and hospitality at JLab during the worshop.



\begin{thebibliography}{}
   
\bibitem{deMelo2002}
J.~P.~B.~C. de Melo, 
T. Frederico, E. Pace and G. Salm\`e, 
Nucl. Phys. {\bf A707}, 399 (2002); ibid. Braz. J. Phys. {\bf 33}, 301 (2003).
 
\bibitem{deMelo2014}
J.~P.~B.~C.~de Melo, K.~Tsushima, B.~El-Bennich, E.~Rojas and T.~Frederico, 
  Phys.\ Rev.\ C {\bf 90}, no. 3, 035201 (2014).
  
\bibitem{Yabusaki2015}
G.~H.~S.~Yabusaki, I.~Ahmed, M.~A.~Paracha, J.~P.~B.~C.~de Melo and B.~El-Bennich,
Phys.Rev. {\bf D92} 3, 034017 (2015).
  
\bibitem{Nam2012v1}Seung-il~Nam and C.~W.~Kao,  Phys.\ Rev.\ D {\bf 85}, 094023 (2012)

\bibitem{Nam2012v2}Seung-il Nam, 
Phys.\ Rev.\ D {\bf 86} (2012) 074005.


\bibitem{Bira1999}
U.~van Kolck,~Prog.Part.Nucl.Phys. 43 (1999) 337-418


\bibitem{Frederico1994}T. Frederico and 
G.~A. Miller,~Phys.~Rev.~{\bf D50},~210~(1994).




  

   
\bibitem{PDG}K.A. Olive et al. (Particle Data Group), Chin. Phys. C, 38, 
090001 (2014) and 2015 update.

\bibitem{Choi2007}
Ho-Meoyng Choi and Chueng-Ryong Ji,~Phys.Rev. {\bf D75} (2007) 034019.


\bibitem{Nico2015}
N.G. Stefanis,~S.V. Mikhailov and 
A.V. Pimikov,~Few Body Syst. {\bf 56} (2015) 6-9, 295.


\bibitem{deMelo1999} 
J.~P.~B.~C. de Melo, H.~W.~L.~Naus and T.~Frederico, Phys.~Rev.~{\bf C59}, 2278 (1999). 

\bibitem{Otoniel2012} Edson O. da Silva, J. P. B. C. de Melo, 
Bruno El-Bennich, and Victo S. Filho,
Phys.~Phys.~Rev.~{\bf C 86}, 038202 (2012).

\bibitem{Bakker2001}B.~L.~G.~Bakker, H.-M.~Choi,~C.-R.~Ji,  Phys.~Rev. D 
{\bf 63},~074014~(2001).
  
\bibitem{Suisso2002}
E.~F.~Suisso, J.~P.~B.~C.~de~Melo,~T.~Frederico,~Phys.Rev. D65 (2002) 094009.
  
\bibitem{Lepage1980}
G.~P.~Lepage and S.~J.~Brodsky,~Phys.~Rev.~{\bf D22}~(1980)~2157. 
 
 \end{thebibliography}
\end{document}